\documentclass[mathleft]{an}
\usepackage{graphicx}
\usepackage{times}
\usepackage{natbib}
\bibpunct{(}{)}{;}{a}{}{,}

\begin{document}

\Pagespan{1}{4}
\Yearpublication{2010}%
\Yearsubmission{2010}%
\Month{00}%
\Volume{000}%
\Issue{00}%
\DOI{This.is/not.aDOI}%

\title{Asteroseismology of Solar-type stars with Kepler II: Stellar Modeling}

\author{T.S.~Metcalfe\inst{1} \and M.J.P.F.G.~Monteiro\inst{2} \and 
M.J.~Thompson\inst{3} \and W.J.~Chaplin\inst{4} \and S.~Basu\inst{5} \and 
A.~Bonanno\inst{6} \and M.P.~Di~Mauro\inst{7} \and G.~Do\u{g}an\inst{8} 
\and P.~Eggenberger\inst{9} \and C.~Karoff\inst{4,8} \and 
D.~Stello\inst{10} \and KASC WG1\thanks{The Kepler Asteroseismic Science 
Consortium Working Group 1 also includes: Appourchaux, T., Elsworth, Y., 
Garcia, R.A., Houdek, G., Molenda-\.{Z}akowicz, J., Brown, T.M., 
Christensen-Dalsgaard,~J., Gilliland, R.L., Kjeldsen, H., Borucki, W.J., 
Koch, D., Jenkins, J.M., Ballot, J., Bazot, M., Bedding, T.R., Benomar, 
O., Brandao, I.M., Bruntt, H., Campante, T.L., Creevey, O.L., Dreizler, 
S., Esch, L., Fletcher, S. T., Frandsen, S., Gai, N., Gaulme, P., 
Handberg, R., Hekker, S., Howe, R., Huber, D., Korzennik, S.G., Lebrun, 
J.C., Leccia, S., Martic, M., Mathur, S., Mosser, B., New, R., Quirion, 
P.-O., Regulo, C., Roxburgh, I.W., Salabert, D., Schou, J., Sousa, S.G., 
Verner, G.A., Arentoft, T., Barban, C., Belkacem, K., Benatti, S., Biazzo, 
K., Boumier, P., Bradley, P.A., Broomhall, A.-M., Buzasi, D.L., Claudi, 
R.U., Cunha, M.S., D'Antona, F., Deheuvels, S., Derekas, A., Garcia 
Hernandez, A., Giampapa, M.S., Goupil, M.J., Gruberbauer, M., Guzik, J.A., 
Hale, S.J., Ireland, M.J., Kiss, L.L., Kitiashvili, I.N., Kolenberg, K., 
Korhonen, H., Kosovichev, A.G., Kupka, F., Lebreton, Y., Leroy, B., 
Ludwig, H.-G., Mathis, S., Michel, E., Miglio, A., Montalban, J., Moya, 
A., Noels, A., Noyes, R.W., Palle, P. L., Piau, L., Preston, H.L., Roca 
Cortes, T., Roth, M., Sato, K.H., Schmitt, J., Serenelli, A.M., Silva 
Aguirre, V., Stevens, I.R., Suarez, J. C., Suran, M.D., Trampedach, R., 
Turck-Chieze, S., Uytterhoeven, K., Ventura, R \& Wilson, P.A.}}

\titlerunning{Asteroseismology of Solar-type Stars with Kepler}
\authorrunning{T.S.~Metcalfe et al.}
\institute{
High Altitude Observatory, National Center for Atmospheric Research, 
P.O. Box 3000, Boulder Colorado 80307 USA
\and 
Centro de Astrof\'{\i}sica and DFA-Faculdade de Ci\^encias, Universidade 
do Porto, Portugal
\and 
School of Mathematics and Statistics, University of Sheffield, 
Hounsfield Road, Sheffield S3 7RH, UK
\and
School of Physics and Astronomy, University of Birmingham, Edgbaston, 
Birmingham, B15 2TT, UK
\and
Department of Astronomy, Yale University, P.O. Box 208101, New Haven, CT 
06520-8101, USA
\and
INAF Osservatorio Astrofisico di Catania, Via S.Sofia 78, 95123, Catania, 
Italy
\and
INAF-IASF, Istituto di Astrofisica Spaziale e Fisica Cosmica, via del Fosso 
del Cavaliere 100, 00133 Roma, Italy
\and
Department of Physics and Astronomy, Aarhus University, DK-8000 Aarhus C, 
Denmark
\and
Geneva Observatory, University of Geneva, Maillettes 51, 1290, Sauverny, 
Switzerland
\and
Sydney Institute for Astronomy (SIfA), School of Physics, University of 
Sydney, NSW 2006, Australia
}

\received{15 April 2010}
\accepted{30 June 2010}
\publonline{}


\abstract{Observations from the Kepler satellite were recently published 
for three bright G-type stars, which were monitored during the first 
33.5~d of science operations. One of these stars, KIC~11026764, exhibits a 
characteristic pattern of oscillation frequencies suggesting that the star 
has evolved significantly. We have derived initial estimates of the 
properties of KIC~11026764 from the oscillation frequencies observed by 
Kepler, combined with ground-based spectroscopic data. We present 
preliminary results from detailed modeling of this star, employing a 
variety of independent codes and analyses that attempt to match the 
asteroseismic and spectroscopic constraints simultaneously.}

\maketitle


\section{Introduction}

In March 2009, NASA launched the {\it Kepler} satellite---a mission 
designed to find habitable Earth-like planets around distant Sun-like 
stars. The satellite consists of a 0.95-m telescope with an array of 
digital cameras that will monitor the brightness of more than 150,000 
solar-type stars with a few parts-per-million precision for between 4-6 
years \citep{bor10}. Some of these stars are expected to have planetary 
systems, and some of the planets will have orbits such that they 
periodically pass in front of the host star, causing a brief decrease in 
the amount of light recorded by the satellite. The depth of such a {\it 
transit} contains information about the size of the planet relative to the 
size of the host star.

Since we do not generally know the precise size of the host star, the 
mission design includes a revolving selection of 512 stars monitored with 
the higher cadence that is necessary to detect solar-like oscillations, 
allowing us to apply the techniques of asteroseismology 
\citep{jcd07,ack10}. Even a relatively crude analysis of such measurements 
can lead to reliable determinations of stellar radii to help characterize 
the planetary systems discovered by the satellite, and stellar ages to 
reveal how such systems evolve over time. For the asteroseismic targets 
that do not contain planetary companions, these data will allow a uniform 
determination of the physical properties of hundreds of solar-type stars, 
providing a new window on stellar structure and evolution.

Initial results from the Kepler Asteroseismic Investigation were presented 
in \cite{gil10}, while a more detailed analysis of the solar-like 
oscillations detected in several early targets was published by 
\cite{cha10}. The latter paper includes observations of three bright 
(V$\sim$9) G~IV-V stars, which were monitored during the first 33.5~d of 
science operations. One of these stars, KIC~11026764 ($\equiv$ 2MASS 
J19212465+4830532 $\equiv$ BD+48~2882), exhibits a characteristic pattern 
of oscillation frequencies suggesting that the star has evolved 
significantly.

\begin{figure}
\includegraphics[width=83mm]{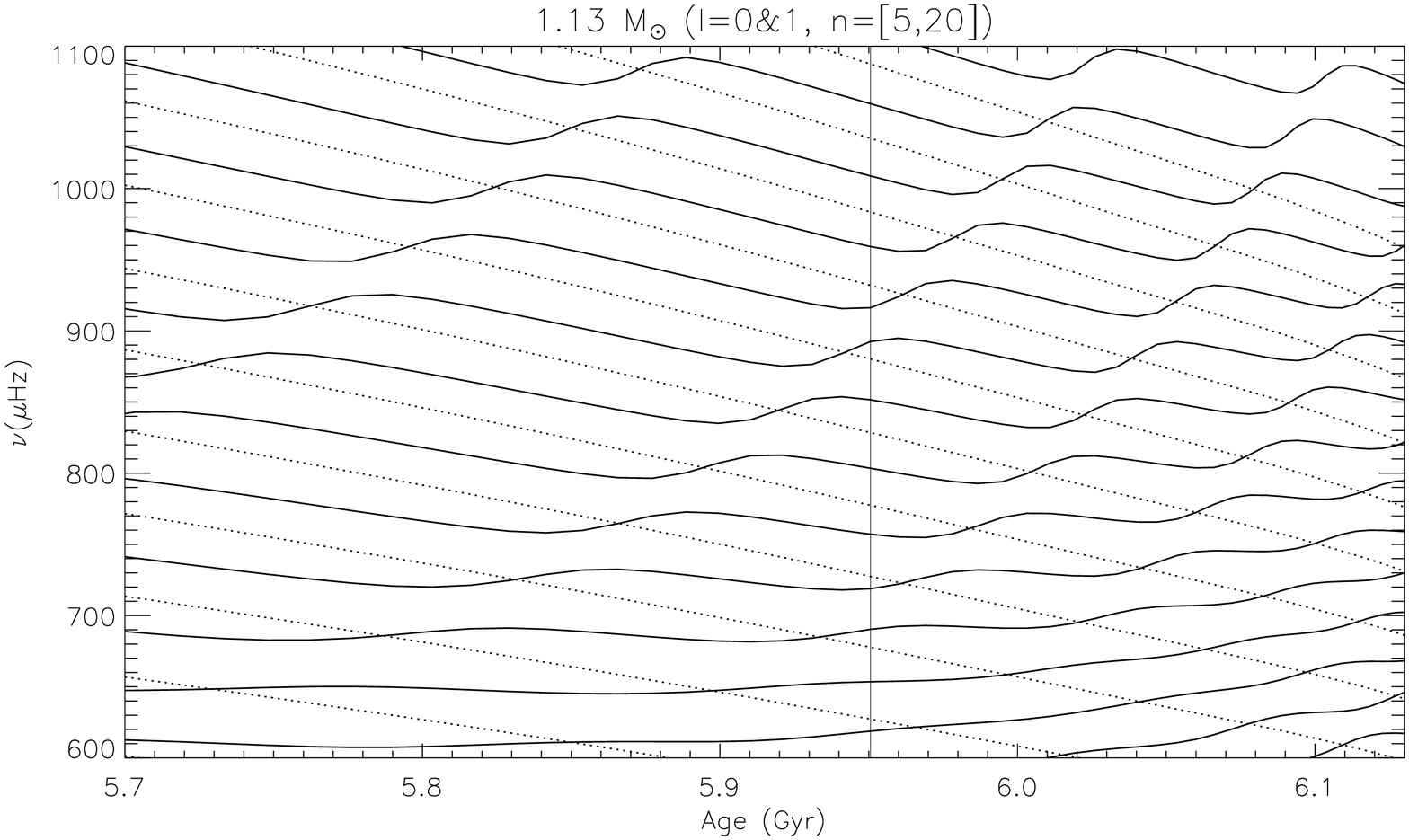}
\caption{Evolution of the $l=0$ (dotted) and $l=1$ (solid) 
oscillation frequencies as a function of age for a representative stellar 
model of KIC~11026764.}
\label{fig1}
\end{figure}

In unevolved stars, the high radial order ($n$) acoustic oscillation modes 
($p$-modes) with a given spherical degree ($l$) are almost evenly spaced 
in frequency. As the star evolves and the envelope expands and cools, the 
$p$-mode frequencies gradually decrease. Meanwhile, as the star becomes 
more centrally condensed, the buoyancy-driven ($g$-mode) oscillations in 
the core shift to higher frequencies. This eventually leads to a range of 
frequencies where the oscillation modes can take on a mixed character, 
behaving like $g$-modes in the core and $p$-modes in the envelope (``mixed 
modes''), with their frequencies shifted as they undergo so-called {\it 
avoided crossings}. This behavior changes very quickly with stellar age, 
and propagates from one radial order to the next as the star continues to 
evolve (see Figure~\ref{fig1}). Consequently, the particular modes that 
deviate significantly from uniform frequency spacing yield a strong 
(though model-dependent) constraint on the age of the star 
\citep[see][]{jcd04}. As noted by \cite{gil10} and \cite{cha10}, the 
dipole ($l=1$) modes observed in KIC~11026764 show the signature of an 
avoided crossing---raising the exciting possibility that detailed modeling 
of this star will ultimately provide a very precise determination of its 
age.

In this paper we derive initial estimates of the stellar properties of 
KIC~11026764 by matching the observed oscillation frequencies from {\it 
Kepler} data and the spectroscopic constraints from ground-based 
observations. The extraction and identification of the oscillation 
frequencies is described by Karoff et al.~(this volume), and the analysis 
of ground-based observations to derive spectroscopic constraints is 
described by Molenda-\.{Z}akowicz et al.~(this volume). We focus on the 
initial results from detailed modeling of this star employing a variety of 
independent codes and analyses, all attempting to match the asteroseismic 
and spectroscopic constraints simultaneously.


\section{Stellar Modeling}

\begin{figure}
\includegraphics[width=83mm]{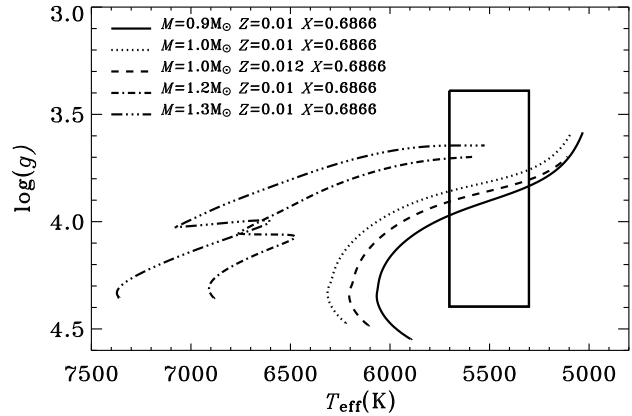}
\caption{The observational error box for KIC~11026764 from the Kepler 
Input Catalog (rectangle) along with several stellar evolution tracks, 
showing that the non-seismic constraints imply either a slightly evolved 
star with a mass comparable to the Sun, or a higher mass star in a more 
advanced stage of evolution.}
\label{fig2}
\end{figure}

Traditional stellar modeling in the absence of asteroseismic information 
involves matching, as closely as possible, the non-seismic constraints in 
a classical Hertzsprung-Russell (H-R) diagram. A spectroscopic 
determination of [Fe/H] can be used to fix the composition of the stellar 
models, and evolution tracks are then typically compared to the available 
constraints on $T_{\rm eff}$ and $\log(g)$ from photometry and 
spectroscopy. The ambiguity of such a comparison is illustrated in 
Figure~\ref{fig2}, which shows the observational error box for 
KIC~11026764 from the Kepler Input Catalog (KIC) along with several 
stellar evolution tracks. Evidently, the non-seismic constraints imply 
either a slightly evolved star with a mass comparable to the Sun, or a 
higher mass star in a more advanced stage of evolution.

For stars that exhibit solar-like oscillations, an estimate of the average 
large and small frequency spacing provides a complementary set of data 
well suited to constraining the stellar properties \citep{jcd93,mct00}. In 
faint KASC survey targets---where lower signal-to-noise ratios make it 
difficult to extract robust estimates of individual frequencies---the 
average spacings will be the primary seismic data. The signatures of these 
spacings are quite amenable to extraction, owing to their near-regularity. 
When the data are sufficient to allow a robust estimation of individual 
frequencies---as is the case for KIC 11026764---use of those frequencies 
increases the information content provided by the seismic data 
\citep[e.g., see][]{rv03,maz06,cm07}.

Different teams extracted estimates of the average separations of 
KIC~11026764, with analysis methods based on autocorrelation of either the 
time series or the power spectrum \citep[e.g., 
see][]{cam10,hek10,hub09,kar10,mat10,ma09,rox09}. We found good agreement 
between the different estimates (i.e., at the level of precision of the 
quoted parameter uncertainties). The teams also used peak-fitting 
techniques \citep[like those applied to CoRoT data; see, for 
example,][]{app08} to provide initial estimates of the individual mode 
frequencies to the modeling teams. Complete details of this analysis can 
be found in Karoff et al. (this volume).

Several modeling teams then applied codes to estimate the stellar 
properties of KIC~11026764 using the frequency separations and other 
non-seismic data as input. The results of these initial analyses were used 
as starting points for further modeling, involving comparisons of the 
observed frequencies with those calculated from evolutionary models. The 
frequencies and the frequency separations depend to some extent on the 
detailed physics assumed in the stellar models \citep{met09,mon02}. 
Consequently, a more secure determination of the stellar properties is 
possible when complementary data are known with sufficiently high accuracy 
and precision. Indeed, the potential to test the input physics of models 
of field stars (e.g., convective energy transport, diffusion, opacities, 
etc.) actually requires non-seismic data for the seismic diagnostics to be 
effective \citep[e.g., see][]{cre07}.

The modeling analyses therefore also incorporated non-seismic constraints, 
using $T_{\rm eff}$, $\log(g)$ and [Fe/H] from complementary ground-based 
spectroscopic observations. For KIC~11026764 only the KIC estimates were 
initially available, with large uncertainties of about 200\,K in $T_{\rm 
eff}$, and up to 0.5\,dex in $\log(g)$ and [Fe/H]. We performed several 
cross-checks of the $T_{\rm eff}$ for KIC~11026764 using different 
suggested temperature calibrations of the available 2MASS \citep{cut03} 
$VJHK$ magnitudes \citep{kc02,gal05,mas06}. These tests yielded 
satisfactory agreement with the $T_{\rm eff}$ value from KIC at the level 
of the estimated uncertainties. Preliminary results given by different 
groups on the same ground-based spectra of these stars do however suggest 
that the true, external errors are higher than the quoted errors. Complete 
details of this analysis can be found in Molenda-\.{Z}akowicz et al. (this 
volume).

\section{Initial Results}

Our initial estimates of the properties of KIC~11026764 using several 
independent codes and analyses are presented in Table~\ref{tab1}. Given 
the close relation between the global properties of the stars and their 
oscillation frequencies, these seismically inferred properties are more 
precise, and more accurate, than properties inferred without the seismic 
inputs.

\begin{table*}
\begin{center}
\centering
\caption{Preliminary model-fitting results for KIC~11026764 from several 
different codes.}
\label{tab1}
\begin{tabular}{rcccccccccl}\hline
Code&$M/M_\odot$&Z&X&$\alpha$&t(Gyr)&$T_{\rm eff}$(K)&$L/L_\odot$&$R/R_\odot$&$\log$\,g&Reference\\
\hline
YREC    & 1.20 & 0.0142 & 0.76 & 1.83 & 6.25 & 5500 & 3.53 & 2.07 & 3.88 & \cite{dem08}\\
Catania & 1.10 & 0.0159 & 0.77 & 1.10 & 6.60 & 5357 & 2.97 & 2.00 & 3.87 & \cite{bon02}\\
ASTEC-1 & 1.00 & 0.0100 & 0.69 & 1.88 & 6.51 & 5700 & 3.60 & 1.95 & 3.86 & \cite{jcd08}\\
ASTEC-2 & 1.10 & 0.0125 & 0.72 & 1.70 & 6.35 & 5653 & 3.65 & 2.01 & 3.88 & \cite{jcd08}\\
Geneva  & 1.15 & 0.0150 & 0.71 & 1.80 & 5.80 & 5581 & 3.73 & 2.07 & 3.86 & \cite{egg08}\\
SEEK    & 1.20 & 0.0197 & 0.71 & 1.15 & 5.50 & 5500 & 3.36 & 2.03 & 3.90 & \cite{qui10}\\
RADIUS&$\cdots$&$\cdots$&$\cdots$&$\cdots$&$\cdots$&$\cdots$&$\cdots$&2.03&$\cdots$&\cite{ste09}\\
\hline
\end{tabular}
\end{center}
\end{table*}

\begin{figure}
\includegraphics[width=83mm]{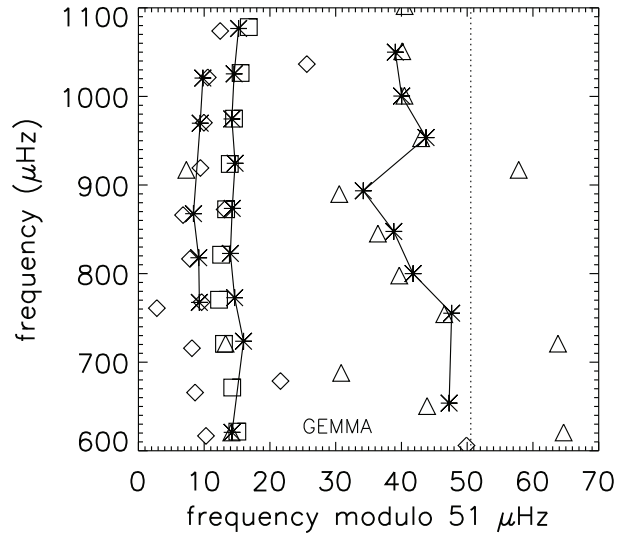}
\caption{An echelle diagram for the observed frequencies of KIC~11026764 
(connected points), where we divide the oscillation spectrum into segments 
of length $\left<\Delta\nu_0\right>$ and plot them against the oscillation 
frequency, along with a representative stellar model (open points) showing 
the $l=0$ (squares), $l=1$ (triangles) and $l=2$ modes 
(diamonds). Note the $l=1$ avoided crossing near 900~$\mu$Hz and the 
overlapping $l=0$ and $l=1$ modes near 620 and 720~$\mu$Hz, which 
were common features in all analyses.}
\label{fig3}
\end{figure}

The precision achieved for KIC~11026764 is about 5\% in the radius, and 
about 10\% in the mass. This star has evolved off the main sequence, and 
is relatively difficult to model. The analysis demonstrates that when 
mixed modes are observed, individual frequencies can provide more 
stringent tests of the modeling than the average frequency spacings alone. 
Initial results from modeling the individual frequencies show that it is 
possible to reproduce the disrupted $l=1$ frequency ridge (see 
Figure~\ref{fig3}), and indicate a stellar age near 6\,Gyr.

These initial results may also help us to interpret the observed seismic 
spectra, allowing additional mode frequencies to be identified securely. 
For example, the prominent mode near 720~$\mu$Hz lies on the $l=0$ ridge, 
yet its appearance suggests a possible alternative explanation. Mixed 
modes exhibit $g$-mode character in the core, so the mode inertia is 
typically much higher than regular $p$-modes at the same frequency 
\citep[see][]{jcd04}. This inertia prevents mixed modes from being as 
strongly damped, leading to narrower peaks in the power spectrum. The 
modeling points strongly to the observed power being predominantly from an 
$l=1$ mode that has been shifted so far in frequency that it lies on top 
of an $l=0$ mode. This was a generic feature of several independent 
analyses using different stellar evolution codes, lending further 
credibility to this interpretation.

In summary, KIC~11026764 is clearly an excellent candidate for long-term 
observations by {\it Kepler}. With up to 1\,yr of data we would expect to 
measure the depth of the near-surface convection zone, and the signatures 
of near-surface ionization of He \citep{ver06}. It should also be possible 
to constrain the rotational frequency splittings. With 2\,yr of data and 
more, we could also begin to constrain any long-term changes to the 
frequencies and other mode parameters due to stellar-cycle effects 
\citep{kar09}. More detailed modeling will also allow us to characterize 
the functional form of any required near-surface corrections to the model 
frequencies \citep[see][]{kje08}. Considering that KIC~11026764 is just 
one of the thousands of solar-type stars that have been observed during 
the survey phase of {\it Kepler}, the future of asteroseismology looks 
very bright.

\acknowledgements 
This work was supported in part by NASA grant NNX09AE59G. The National 
Center for Atmospheric Research is a federally funded research and 
development center sponsored by the U.S.~National Science Foundation.
GD and CK acknowledge financial support from the Danish Natural Science 
Research Council. Funding for the Kepler mission is provided by NASA's
Science Mission Directorate. We thank the entire Kepler team for the 
development and operations of this outstanding mission.


\end{document}